\documentclass[Physsubmission, Phys]{SciPost}

\hypersetup{
    colorlinks,
    linkcolor={red!50!black},
    citecolor={blue!50!black},
    urlcolor={blue!80!black}
}

\usepackage[bitstream-charter]{mathdesign}
\usepackage{float}
\usepackage{txfonts}
\usepackage{wasysym}
\usepackage{graphicx}
\usepackage{color}

\usepackage[normalem]{ulem}  

\begin{document}

\begin{center}{\Large \textbf{
Predictions for neutron stars from holographic nuclear matter \\
}}\end{center}

\begin{center}
Nicolas Kovensky\textsuperscript{1},
Aaron Poole\textsuperscript{2} and
Andreas Schmitt\textsuperscript{2$\star$}
\end{center}

\begin{center}
{\bf 1} Institut de Physique Th\'eorique, Universit\'e Paris Saclay, CEA, CNRS, Orme des Merisiers, 91191 Gif-sur-Yvette CEDEX, France.
\\
{\bf 2} Mathematical Sciences and STAG Research Centre, University of Southampton, Southampton SO17 1BJ, United Kingdom
\\
* a.schmitt@soton.ac.uk
\end{center}

\begin{center}
20 December, 2021
\end{center}


\definecolor{palegray}{gray}{0.95}
\begin{center}
\colorbox{palegray}{
  \begin{minipage}{0.95\textwidth}
    \begin{center}
    {\it  XXXIII International (ONLINE) Workshop on High Energy Physics \\“Hard Problems of Hadron Physics:  Non-Perturbative QCD \& Related Quests”}\\
    {\it November 8-12, 2021} \\
    \end{center}
  \end{minipage}
}
\end{center}

\section*{Abstract}
{\bf
We discuss masses, radii, and tidal deformabilities of neutron stars constructed from the  holographic Witten-Sakai-Sugimoto model. Using the same model for crust and core of the star, we combine our theoretical results  with the latest astrophysical data, thus deriving more stringent constraints than given by the data alone. For instance, our calculation predicts -- independent of the model parameters -- an upper limit for the maximal mass of the star of about 2.46 solar masses and a lower limit of the (dimensionless) tidal deformability of a 1.4-solar-mass star of about 277.
}


\section{Introduction}
\label{sec:intro}

The gauge-gravity duality \cite{Maldacena:1997re,Witten:1998zw} is a powerful non-perturbative tool to understand strongly coupled gauge theories. Based on the holographic principle, it is employed to obtain otherwise inaccessible strong-coupling results from classical gravity calculations in higher dimensions. Here we use a certain realization of the gauge-gravity duality, the Witten-Sakai-Sugimoto model \cite{Witten:1998zw,Sakai:2004cn,Sakai:2005yt}, to describe  cold and dense matter at baryon and isospin densities relevant for neutron stars.

Neutron stars present a unique laboratory for matter at large, but not asymptotically large, densities, where first-principle calculations within Quantum Chromodynamics (QCD) are too difficult within currently available techniques (for recent progress on the lattice see for instance Refs.\ \cite{Philipsen:2019qqm,Ito:2020mys,Borsanyi:2021sxv}). The interior of neutron stars is therefore often studied with the help of phenomenological models, effective field theories, or extrapolations of perturbative results, and the resulting thermodynamic and transport properties can be linked to astrophysical observables \cite{Schmitt:2010pn,Schmitt:2017efp}. In recent years, an increasing amount of astrophysical data has become available, for instance through the detection of gravitational waves from neutron star mergers  \cite{LIGOScientific:2018hze,LIGOScientific:2020zkf} and through the NICER mission \cite{Riley:2019yda,Miller:2019cac,Riley:2021pdl,Miller:2021qha}. We shall combine the inferred estimates for mass, radius, and tidal deformability of neutron stars
with our holographic calculation. 

The Witten-Sakai-Sugimoto model is the holographic top-down approach -- based on type-IIA string theory -- that is  closest to real-world QCD. It accounts for chiral and deconfinement phase transitions, and several candidate phases at high densities can be realized, including a holographic version of quarkyonic matter \cite{Kovensky:2020xif}, which, however, 
tends to appear at densities larger than expected in the cores of neutron stars. Here we restrict ourselves to nuclear matter  with two flavors, $N_f=2$, i.e., hybrid stars with a quark matter core or a quarkyonic core will not be discussed.  We employ the holographic results for the core of the star and, within a simple approximation, for the crust as well, such that the crust-core transition is determined dynamically. In this regard, our study goes beyond previous holographic approaches to neutron stars \cite{Hoyos:2016zke,Jokela:2018ers,BitaghsirFadafan:2020otb,Jokela:2020piw} and beyond many field-theoretical studies, where the crust is often obtained from a separate approach and assumptions about the crust-core transition have to be added by hand. Different holographic approaches have recently been reviewed and compared in Ref.\ \cite{Hoyos:2021uff}. 

Secs.\ \ref{sec:holo} and \ref{sec:MR} of these proceedings provide a review of the results of Ref.\ \cite{Kovensky:2021kzl}. However, we  significantly enhance these results by combining them more systematically with the astrophysical data, thus extracting novel predictions for mass, radius, and tidal deformability of the star in Sec.\ \ref{sec:predict}.

\section{Holographic approach}
\label{sec:holo}

\subsection{Model and approximations}

We work within the background geometry of the Witten-Sakai-Sugimoto model that corresponds to the confined phase. The background is given by $N_c$ D4-branes, where $N_c$ corresponds to the number of colors in the dual gauge theory. The $N_f$ D8- and $\overline{\rm D8}$-branes, added to describe left- and right-handed fermions \cite{Sakai:2004cn,Sakai:2005yt}, are assumed to be maximally separated asymptotically in a compact extra dimension with radius $M_{\rm KK}^{-1}$, such that their embedding in the bulk follows geodesics. In this version of the model, there are only two parameters: the 't Hooft coupling $\lambda$ and the Kaluza-Klein mass $M_{\rm KK}$, and we shall discuss our results in this parameter space systematically (setting $N_c=3$, $N_f=2$). We approximate the Dirac-Born-Infeld part of the 
gauge field action on the flavor branes by the Yang-Mills action. The Chern-Simons part of the action 
is crucial to implement nonzero baryon number, and we shall introduce baryonic matter within the "homogeneous ansatz" \cite{Rozali:2007rx,Li:2015uea}. In this ansatz, the spatial components of the non-abelian part of the gauge field are assumed to depend only on the holographic (radial) coordinate, not on the spatial ones. In contrast to an instantonic approach \cite{Preis:2011sp,BitaghsirFadafan:2018uzs}, this approximation is expected to be justified at sufficiently large baryon densities. All our results are valid at zero temperature. 
For the details of the theoretical setup see Ref.\ \cite{Kovensky:2021ddl}, where it was shown how to include an isospin chemical potential in the presence of baryonic matter. This is crucial for the description of realistic neutron star matter. In Ref.\ \cite{Kovensky:2021ddl},  pion condensation was also included and its competition and coexistence with nuclear matter in the phase diagram was investigated. Here we ignore pion condensation for simplicity.  We also neglect the current quark masses (since we only discuss non-strange matter, this is a very good approximation in our context), whose 
effect on the phase structure in the present model was studied in Refs.\ \cite{Kovensky:2019bih,Kovensky:2020xif}. 
The holographic nuclear matter thus constructed shares several properties with real-world nuclear matter, such as a first-order baryon onset of isospin-symmetric nuclear matter. A caveat of the approximation arises due to the semi-classical large-$N_c$ treatment of the baryons. In this treatment, the isospin spectrum is continuous, and neutron and proton states are not explicitly present. While we can still identify the two isospin components with the neutron and the proton, the continuous spectrum is responsible for a symmetry energy at saturation density that is an order of magnitude larger than in the real world. We shall see momentarily that this results in a very large proton fraction of our neutron star matter.  

\subsection{Holographic crust} 

We combine our holographic nuclear matter with a lepton gas made of electrons and muons. Requiring equilibrium with respect to the electroweak interaction and local charge neutrality defines the spatially homogeneous matter in the neutron star core. We also allow for a mixed phase of nuclear matter (plus leptons) and a lepton gas. For the construction of this mixed phase -- which forms the crust of the neutron star -- we require global charge neutrality and assume the interfaces between the two phases to be sharp surfaces. This assumption requires us to introduce the surface tension of nuclear matter $\Sigma$ as an additional external parameter. We assume $\Sigma$ to be constant throughout the crust and will mostly use $\Sigma=1\, {\rm MeV}/{\rm fm}^2$, which is a realistic value for symmetric nuclear matter at saturation density
(roughly the density of our nuclear matter clusters in the crust, up to the crust-core transition). 
Moreover, we employ the Wigner-Seitz approximation and restrict ourselves to the spherical geometry, i.e., we only consider spherical bubbles of nuclear matter
(with dynamically determined size and composition)  immersed in a lepton gas, as expected for the outer crust of the star. We do not construct a mixed phase of nuclear matter with  pure neutron matter, as  expected 
for the inner crust. After these simplifications, the holographic equations of motion together with the neutrality and beta-equilibrium conditions yield the preferred phase for any neutron chemical potential $\mu_n$ fully dynamically. 

\begin{figure}[t]
\centering
\includegraphics[width=0.49\textwidth]{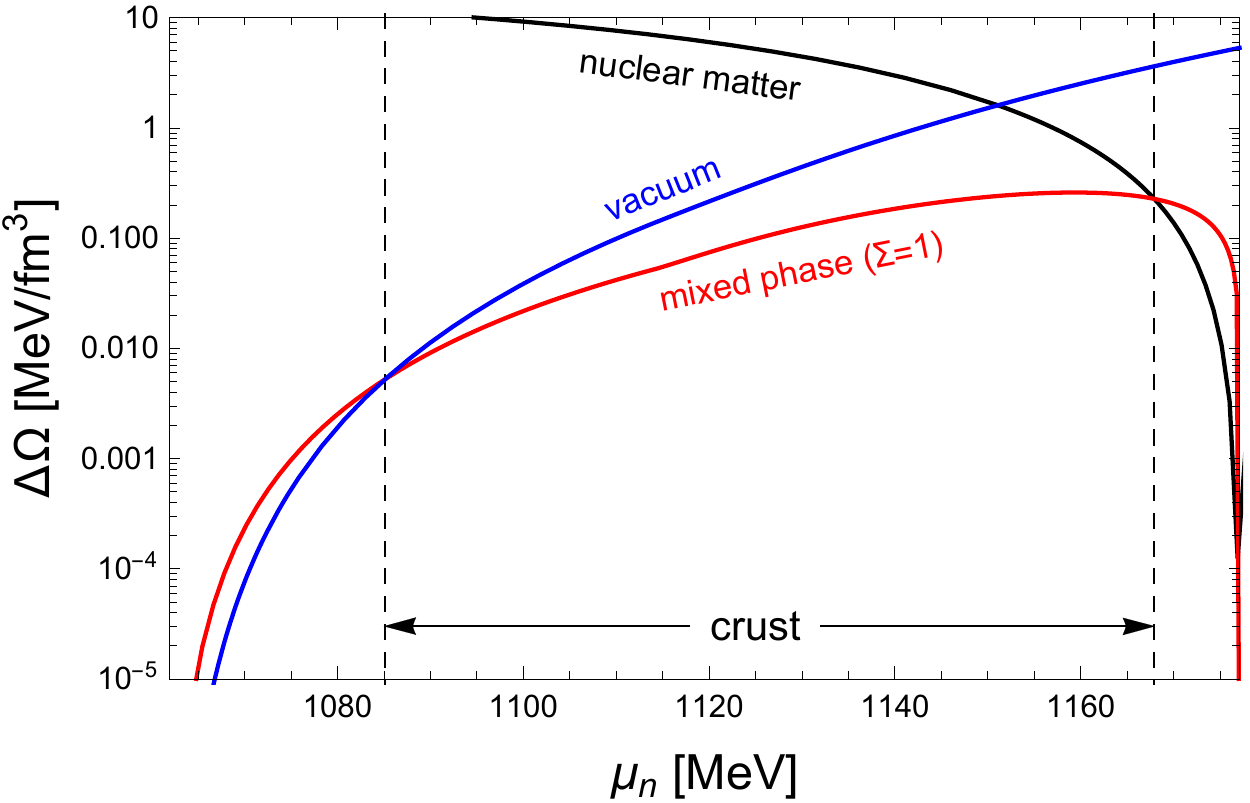}
\includegraphics[width=0.49\textwidth]{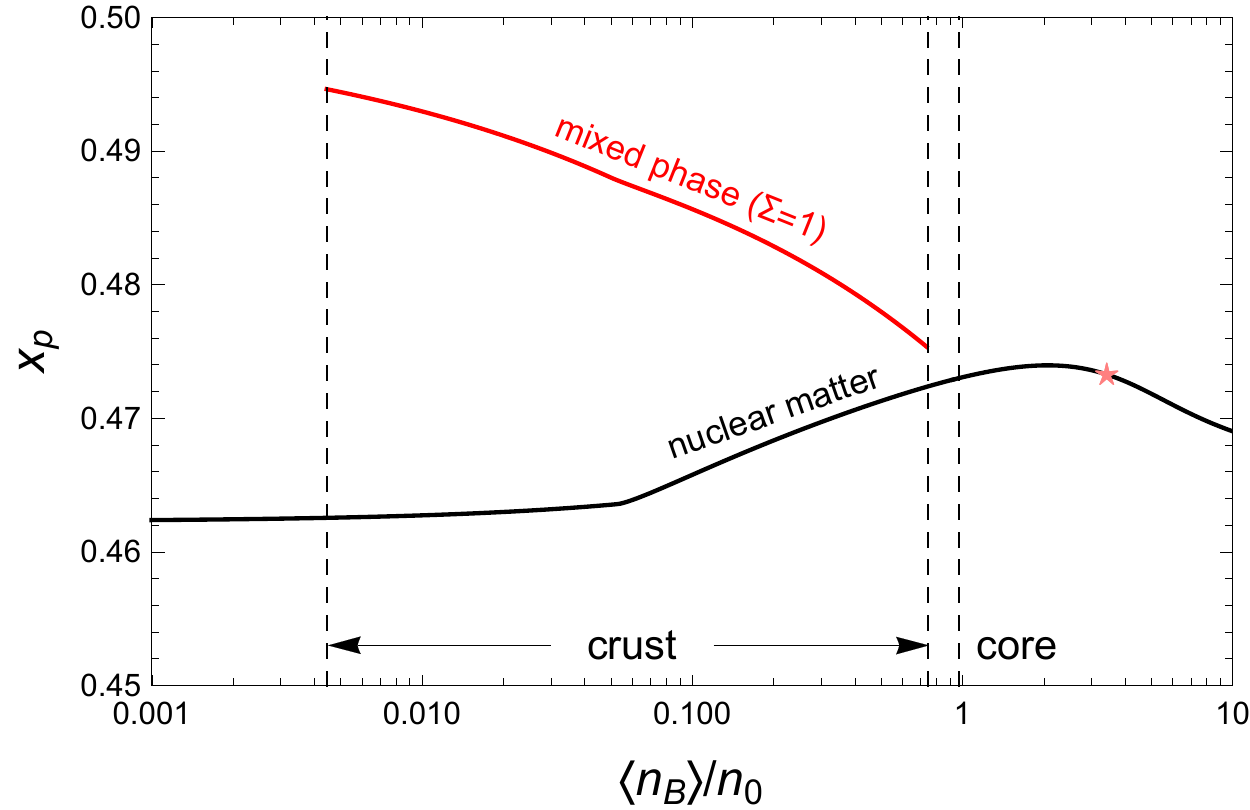}
\caption{{\it Left panel:} Free energy densities relative to the mixed phase without Coulomb and surface effects as a function of the neutron chemical potential. The surface tension is set to $\Sigma = 1\, {\rm MeV}/{\rm fm}^2$. {\it Right panel:} Corresponding proton fraction as a function of the spatially averaged baryon density, normalized by the saturation density of nuclear matter. The star indicates the density and proton fraction in the center of the most massive star. For both panels, $\lambda=10$, $M_{\rm KK} = 949\, {\rm MeV}$, resulting in a saturation density $n_0\simeq 0.21\, {\rm fm}^{-3}$, somewhat larger than in QCD.}
\label{fig:xp}
\end{figure}

We show the results for a certain parameter set in Fig.\ \ref{fig:xp}. The left panel compares the free energy densities of the vacuum, homogeneous nuclear matter, and the mixed phase including Coulomb and surface effects to the free energy density of the mixed phase without Coulomb and surface effects. We read off the transitions between the vacuum and the mixed phase (this will correspond to the surface of the star) and the transition from the mixed phase to homogeneous nuclear matter (crust-core transition). The right panel shows the corresponding proton fraction $x_p = n_p/n_B$, where $n_p$ and $n_B$ are proton and baryon number densities, respectively. We see that our nuclear matter evolves from almost symmetric nuclear matter to more asymmetric matter as we approach the crust-core transition. Then, in the core of the star, the proton fraction rises until at ultra-high densities it decreases again. We also see that the values for $x_p$ are close to 0.5 throughout. This indicates that there is a large energy cost associated with creating isospin-asymmetric matter, which can be attributed to the large-$N_c$ approximation of our approach. Improving the approach to overcome this unrealistic feature is an important step for future work.

\section{Holographic neutron stars}

\subsection{Mass-radius curves} 
\label{sec:MR}

\begin{figure}[t]
\centering
\includegraphics[width=0.49\textwidth]{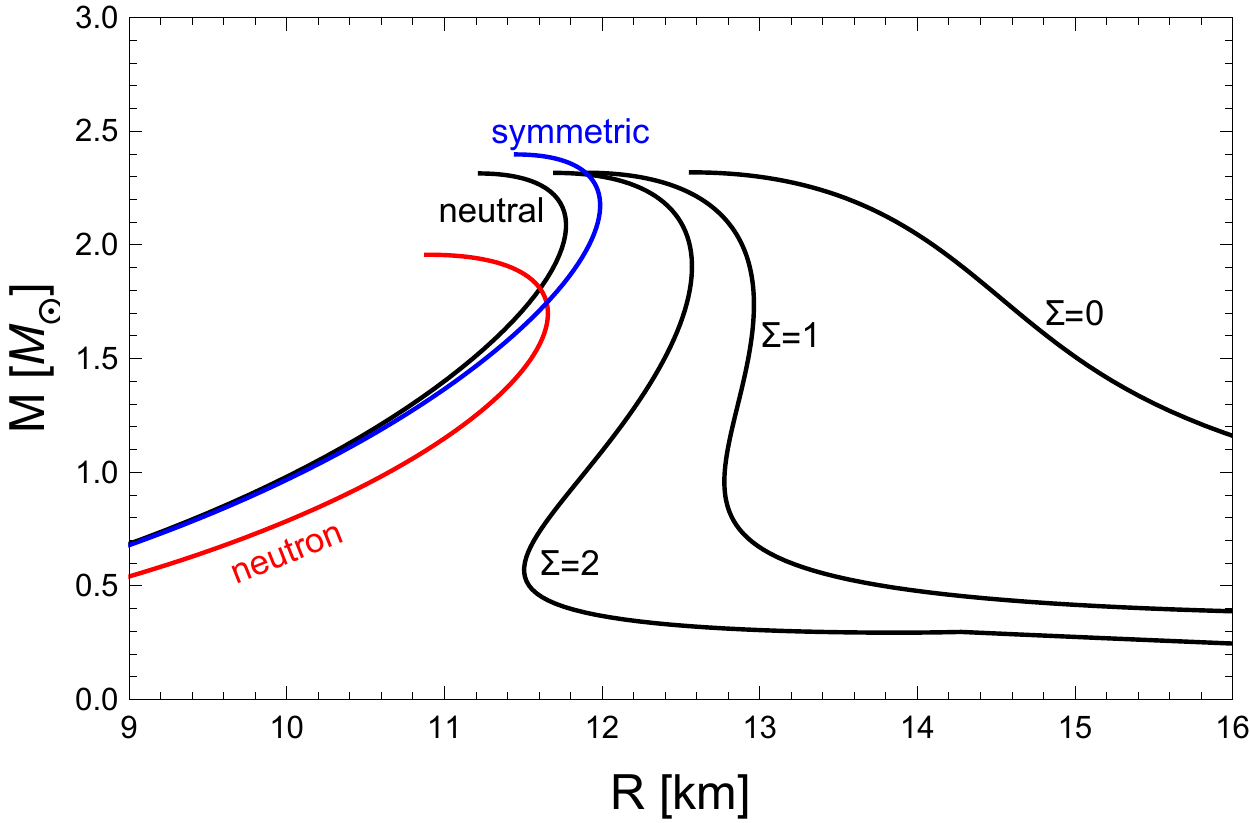}
\includegraphics[width=0.49\textwidth]{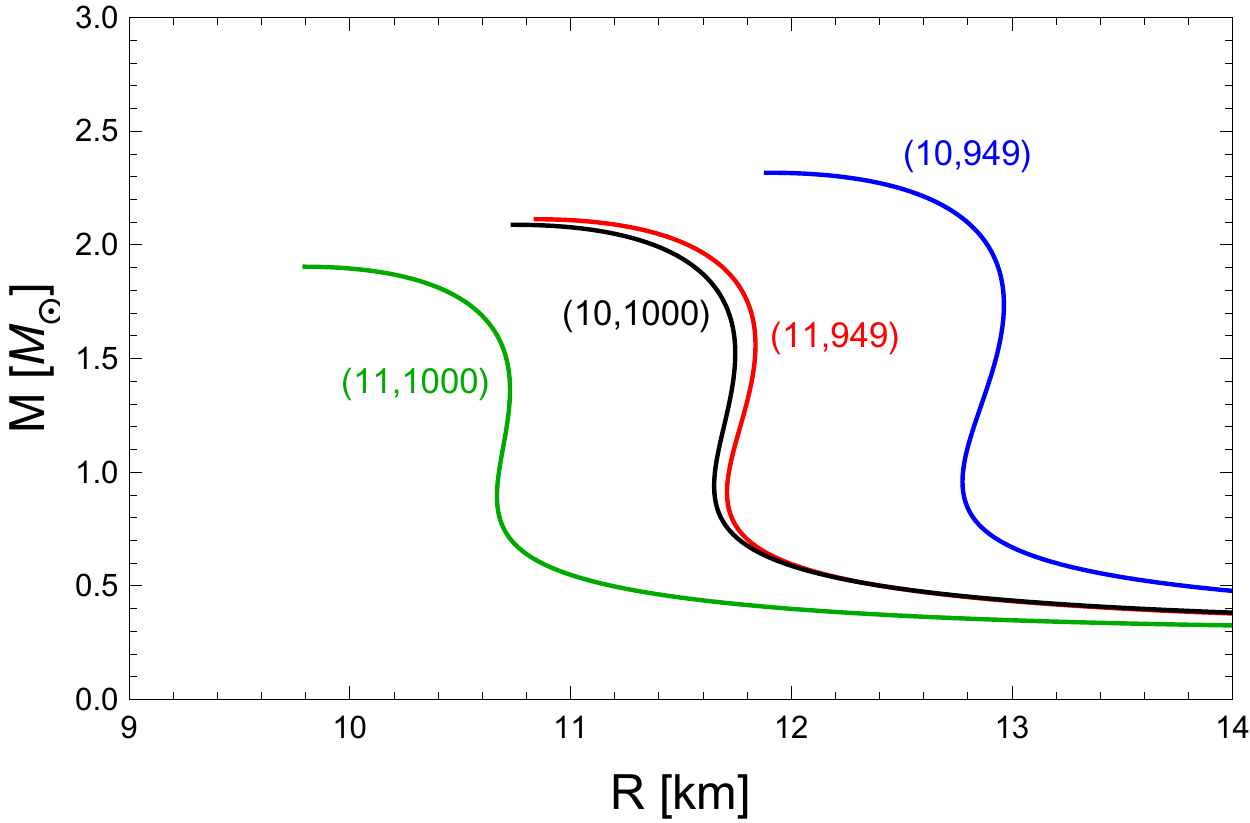}
\caption{{\it Left panel:} Black curves show the effect of the crust, from no crust at all (left) through crust with Coulomb and surface effects (middle, surface tension as labeled, in units of ${\rm MeV}/{\rm fm}^2$) to crust without energy cost (right). For comparison, also the curves for symmetric nuclear matter and pure neutron matter are shown (blue and red, both without crust).
In this panel, $\lambda=10$, $M_{\rm KK}=949$. {\it Right panel:}  Mass-radius curves including the crust with $\Sigma = 1\, {\rm MeV}/{\rm fm}^2$ for different model parameters $\lambda$ and $M_{\rm KK}$ (in MeV), as labeled. All curves end at the maximal mass, beyond which the stars are unstable with respect to radial oscillations.}
\label{fig:MR}
\end{figure}

The holographic calculation laid out in the previous section provides us with all thermodynamic quantities. We can thus straightforwardly compute the equation of state, i.e., the pressure as a function of energy density, including the first-order phase transition at the crust-core boundary, and the corresponding speed of sound. Equation of state and speed of sound are then used as input for the Tolman-Oppenheimer-Volkoff equations (supplemented by an equation for the perturbation of the metric due to tidal deformations), which are solved numerically to extract gravitational mass $M$, radius $R$, and tidal deformability $\Lambda$ for a given central pressure of the star. Varying the central pressure yields mass-radius relations as presented in Fig.\ \ref{fig:MR}. The left panel of this figure shows the effect of the crust and different surface tensions: ignoring the crust leads to very small radii, a crust without Coulomb and surface effects yields very large radii (then the crust is unrealistically large), while Coulomb and surface effects render the effect of the crust more moderate, resulting in radii between the two extremes. The maximal mass is almost unaffected by these changes. The left panel also shows the comparison with pure neutron matter and isospin-symmetric nuclear matter. For the right panel, we have fixed the surface tension and have varied the model parameters $\lambda$ and $M_{\rm KK}$. This panel suggests that realistic "holographic stars" can be obtained. In particular, masses above $2.1\, M_\odot$, where $M_\odot$ is the solar mass, are reached, which is a necessary requirement on account of the observation of the heaviest known neutron star \cite{NANOGrav:2019jur}. We shall confront our results with the other known constraints in the next subsection and see that indeed all known astrophysical constraints can be satisfied (in contrast to the simple pointlike approximation of baryons within the same holographic model \cite{Zhang:2019tqd,Kovensky:2021kzl}).

\begin{figure}[t]
\centering
\includegraphics[width=0.49\textwidth]{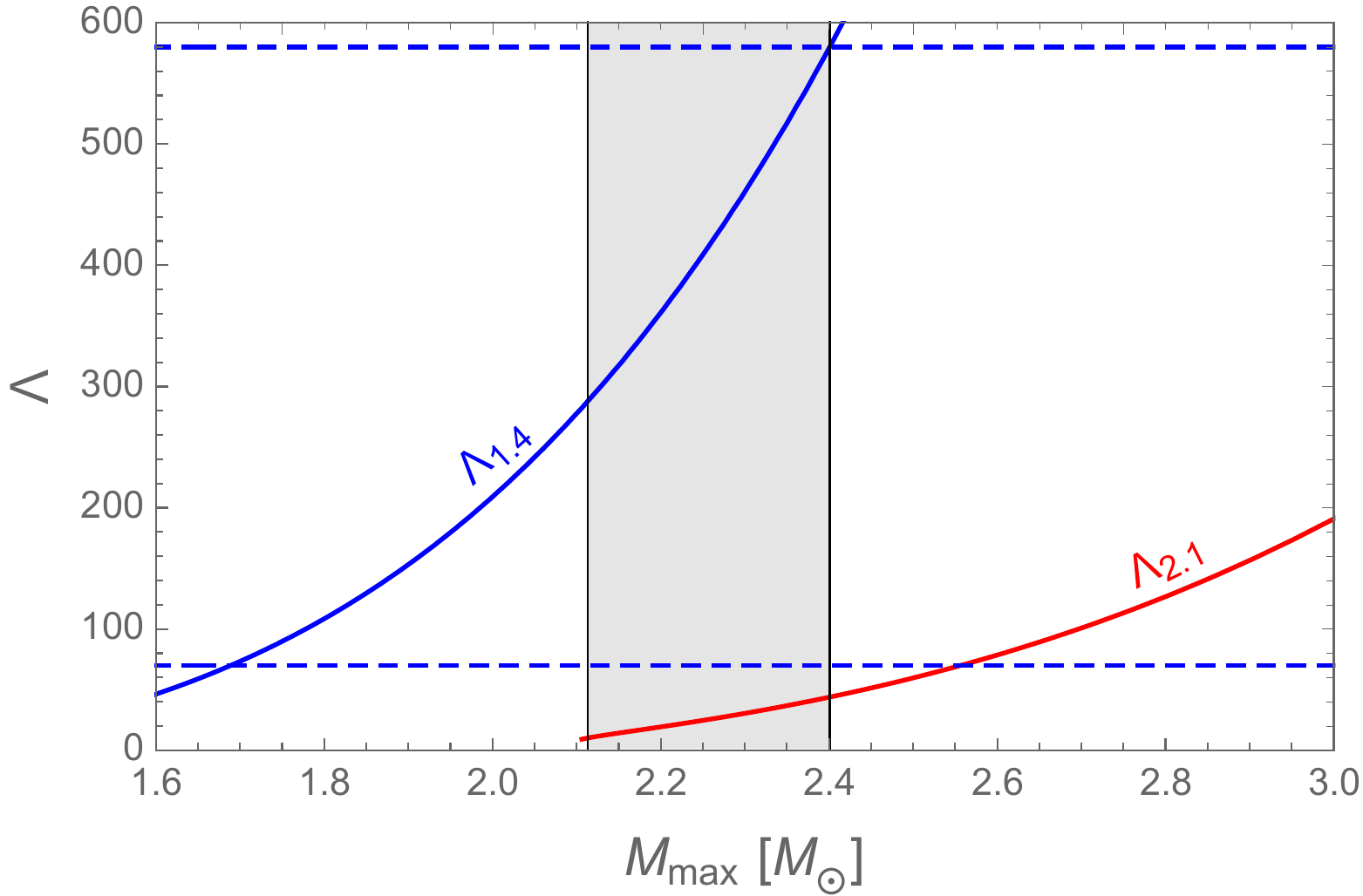}
\includegraphics[width=0.49\textwidth]{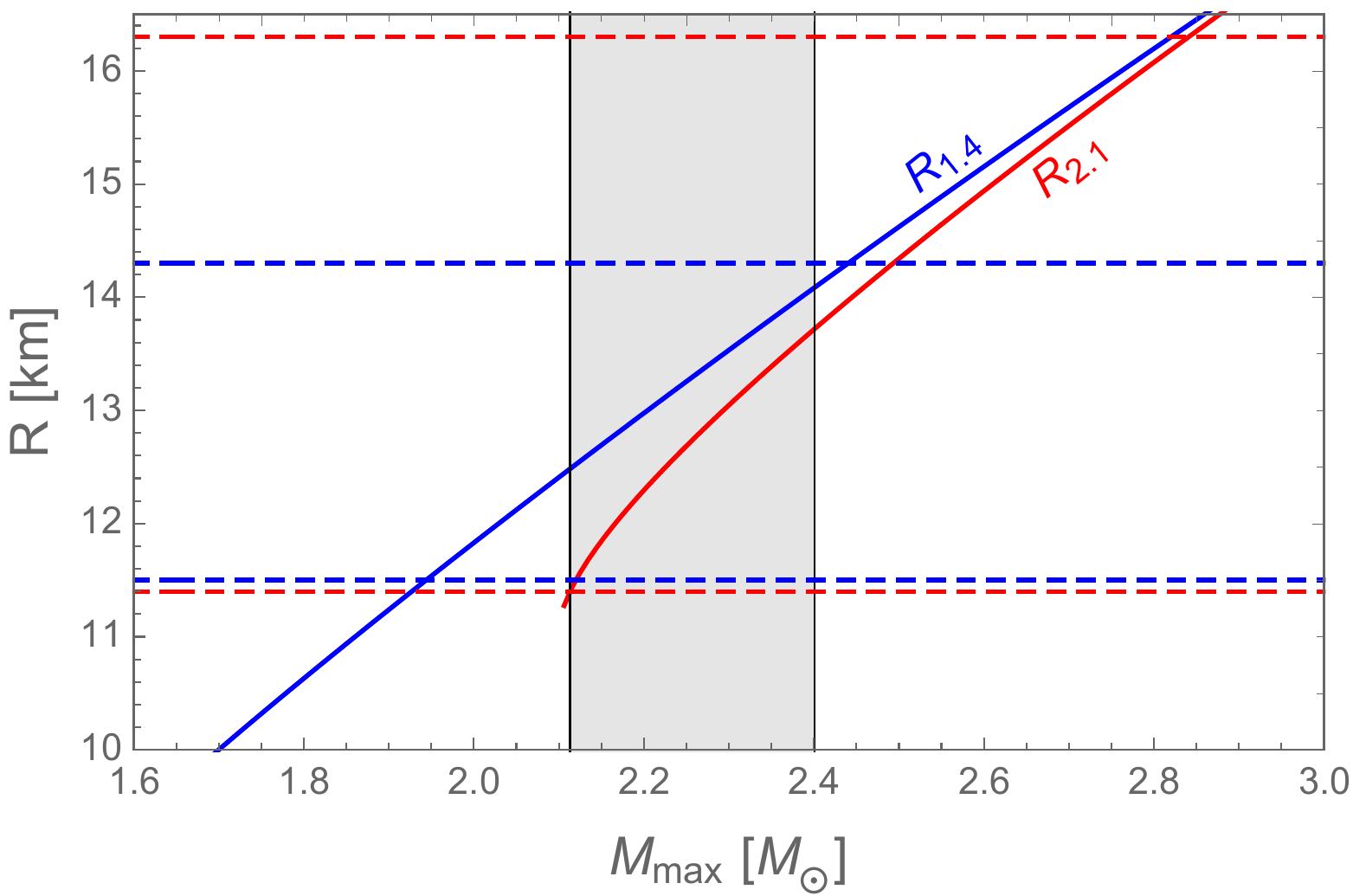}
\caption{Tidal deformability $\Lambda$ and radius $R$ for a 1.4-solar-mass star (blue solid) and a 2.1-solar-mass star (red solid) as a function of the maximal mass $M_{\rm max}$. Here we have fixed the 't Hooft coupling $\lambda = 10$ and the surface tension $\Sigma=1\, {\rm MeV}/{\rm fm}^2$, and different values of $M_{\rm max}$ are obtained by varying the second model parameter $M_{\rm KK}$. The horizontal dashed lines indicate the astrophysical constraints for $\Lambda_{1.4}$, 
$R_{1.4}$, and $R_{2.1}$.
Beyond the shaded region at least one of the constraints is violated. As a consequence, for this particular value of the 't Hooft coupling, we obtain
$2.11\,M_\odot \lesssim M_{\rm max} \lesssim 2.40\,M_\odot$ and new bounds for $\Lambda_{1.4}$, $R_{1.4}$, $\Lambda_{2.1}$, $R_{2.1}$, for instance $288\lesssim \Lambda_{1.4} \lesssim 580$.}
\label{fig:LR}
\end{figure}

\begin{table}[h]
\centering
   \begin{tabular}{|c||c|c|c|}
    \hline
       $\;\;$ fit to $\;\;$  & $\;\;\lambda\;\;$ & $\;\;M_{\rm KK}\;\;$ & Figs.\ \ref{fig:QCD1}, \ref{fig:QCD2}\\
    \hline\hline
        $f_\pi$, $m_\rho$  & $16.63$ & $949\, {\rm MeV}$ & $\CIRCLE$ \\
    \hline
     $\;\;$ $\sigma$, $m_\rho$ $\;\;$  & $\;\;$$12.55$$\;\;$ & $949\, {\rm MeV}$ & $\Diamondblack$\\
    \hline
     $n_0$, $\mu_0$  & $7.09$ & $\;\;$$1000\, {\rm MeV}$$\;\;$ & $\blacksquare$ \\
    \hline     
     \end{tabular}
    \caption{Fits of the model parameters to vacuum properties (pion decay constant and rho meson mass
    \cite{Sakai:2004cn,Sakai:2005yt}, first row,  QCD string tension and rho meson mass \cite{Brunner:2015oqa}, second row), and to nuclear saturation properties (saturation density $n_0=0.153\, {\rm fm}^{-3}$ and onset chemical potential $\mu_0=922.7\, {\rm MeV}$ of symmetric nuclear matter, third row, this work).  
    }
    \label{tab:para}
\end{table}

\subsection{Combining holographic results with astrophysical constraints} \label{sec:predict}

Besides the existence of a $2.1$-solar-mass star, we also consider the constraints for the tidal deformability, $70< \Lambda_{1.4} < 580$ \cite{LIGOScientific:2018hze}, and radius, $11.5\, {\rm km} < R_{1.4} < 14.3\, {\rm km} $ (putting together Refs.\ \cite{Riley:2019yda,Miller:2019cac}), of a (roughly) 1.4-solar-mass star as well as 
the radius, $11.4\, {\rm km} < R_{2.1} < 16.3\, {\rm km} $ (putting together Refs.\ \cite{Riley:2021pdl,Miller:2021qha}), of a (roughly) 
$2.1$-solar-mass star. We demonstrate in Fig.\ \ref{fig:LR} how these constraints can be combined with our holographic results to derive more stringent conditions for mass, radius, and tidal deformability. To obtain this figure, we have fixed the 't Hooft coupling $\lambda$ and calculated the properties of 1.4-solar-mass and 2.1-solar-mass stars and the maximal mass $M_{\rm max}$ for different values of $M_{\rm KK}$. This results in the red and blue solid curves (the curves for the 
$2.1\, M_\odot$ star obviously end where the maximal mass drops below that value). Since the microscopic calculation of homogeneous nuclear matter becomes independent of $M_{\rm KK}$ in the absence of any additional energy scale, we have ignored the muon contribution and set the electron mass to zero here and for the following results. (The surface tension does introduce another energy scale and thus a dependence on $M_{\rm KK}$, but its effect is computed without much numerical effort once the main holographic calculation for a given $\lambda$ is done.)
We now compare the solid curves with the astrophysical constraints, indicated by the horizontal dashed lines. It turns out that the strongest constraint for the upper limit of $M_{\rm max}$ is the upper limit of $\Lambda_{1.4}$ while the strongest constraint for the lower limit of $M_{\rm max}$ is the lower limit of $R_{2.1}$. This gives the two vertical lines, which define the shaded region. This region, in turn, gives new upper limits for $R_{1.4}$, $R_{2.1}$, and new lower limits for $R_{1.4}$, $\Lambda_{1.4}$  (as well as upper and lower limits for $\Lambda_{2.1}$, for which no constraints are known).

\begin{figure}[t]
\centering
\includegraphics[width=0.49\textwidth]{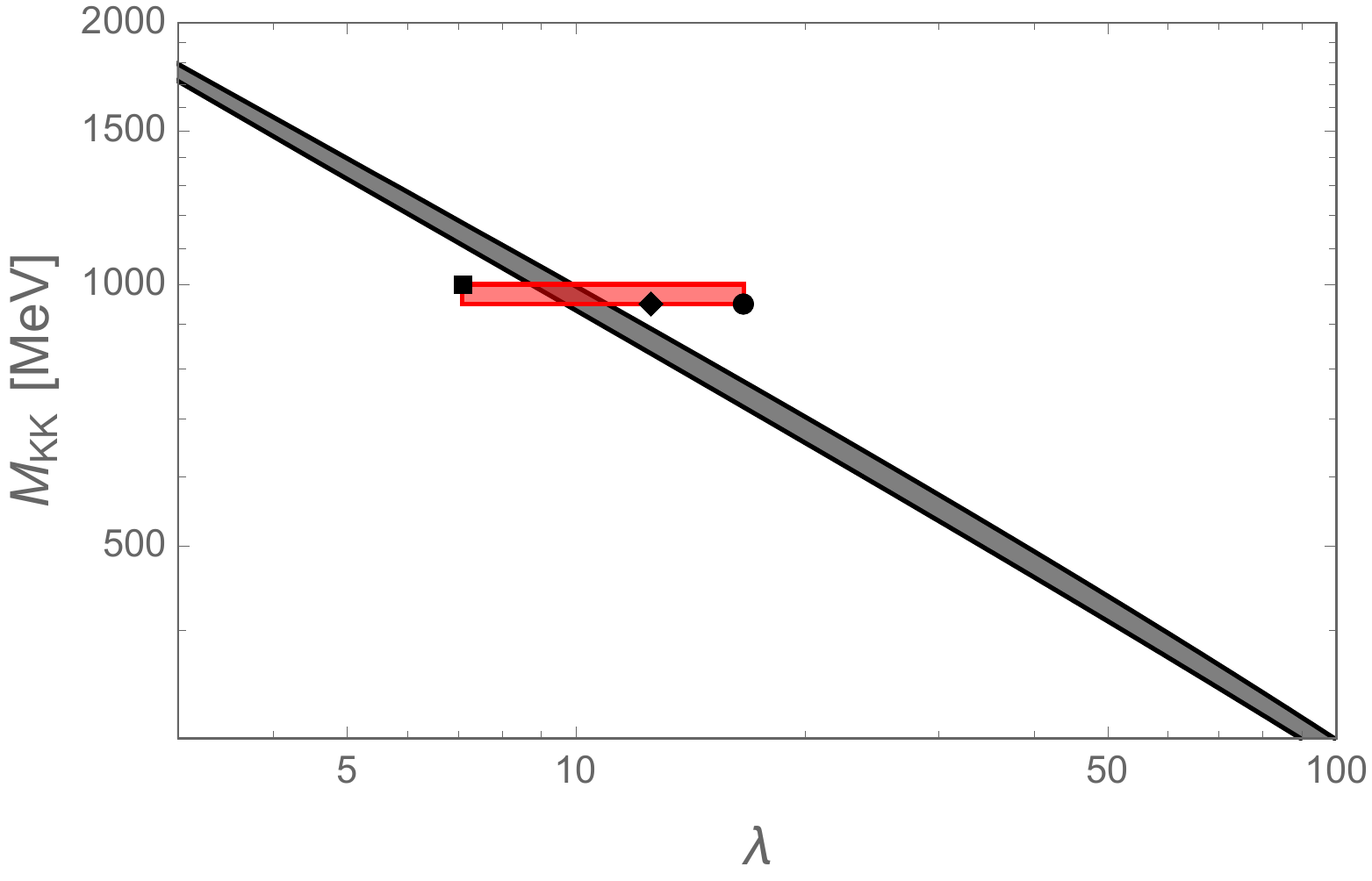}
\includegraphics[width=0.49\textwidth]{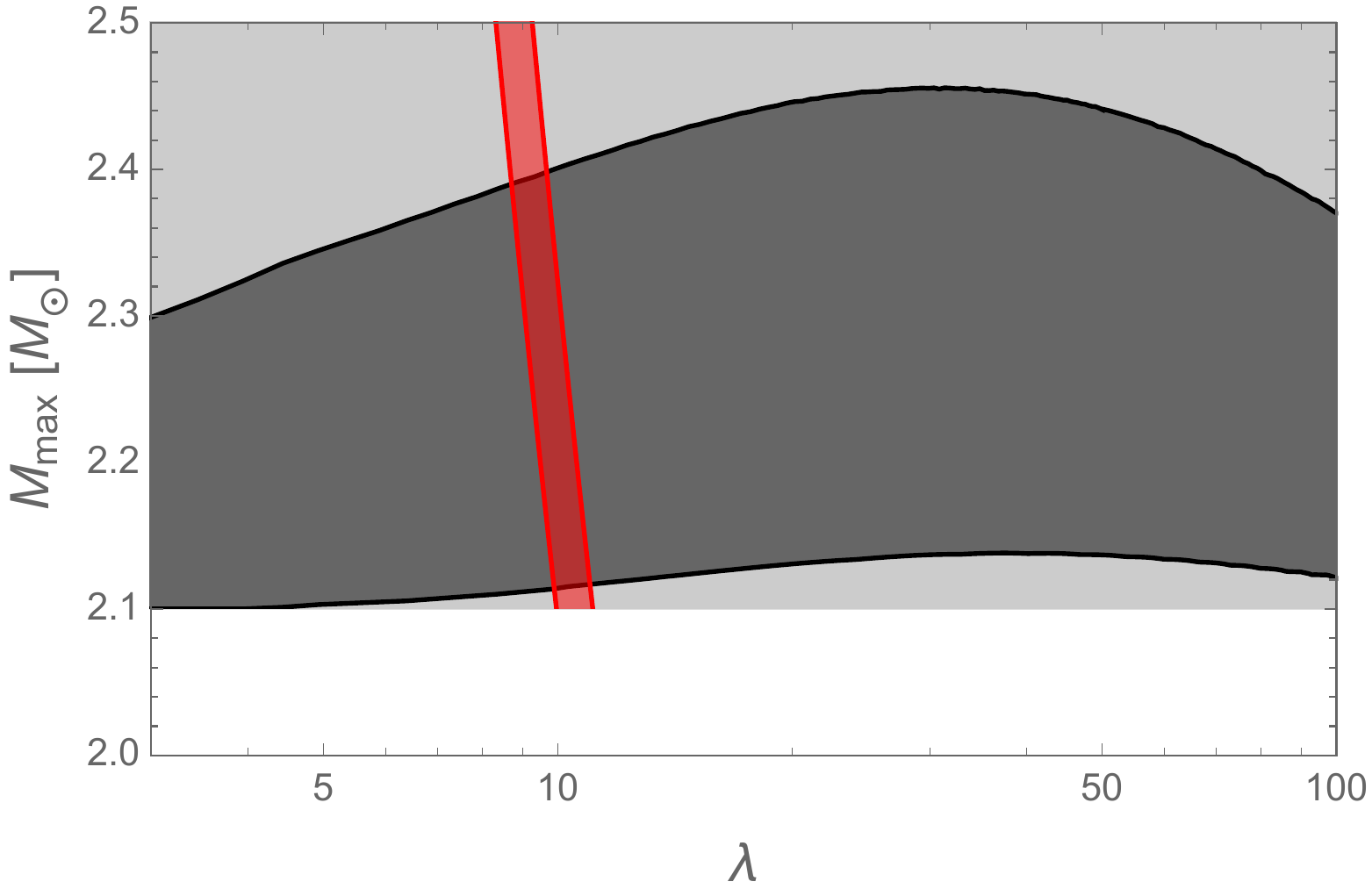}
\caption{{\it Left panel:} Allowed range according to the astrophysical constraints in the $\lambda$-$M_{\rm KK}$ plane (doubly logarithmic), obtained by applying the construction of Fig.\ \ref{fig:LR} for each $\lambda$. The three symbols mark the parameter pairs from the QCD fits of table \ref{tab:para}, and we use them to define a "QCD window" (red). {\it Right panel:} Constraints for the maximal mass  of the neutron star as a function of the 't Hooft coupling $\lambda$. The light gray band gives the constraint from astrophysical data. The dark gray band and the red band arise from applying the constraints of the left panel.
}
\label{fig:QCD1}
\end{figure}
\begin{figure}[H]
\centering
\includegraphics[width=0.49\textwidth]{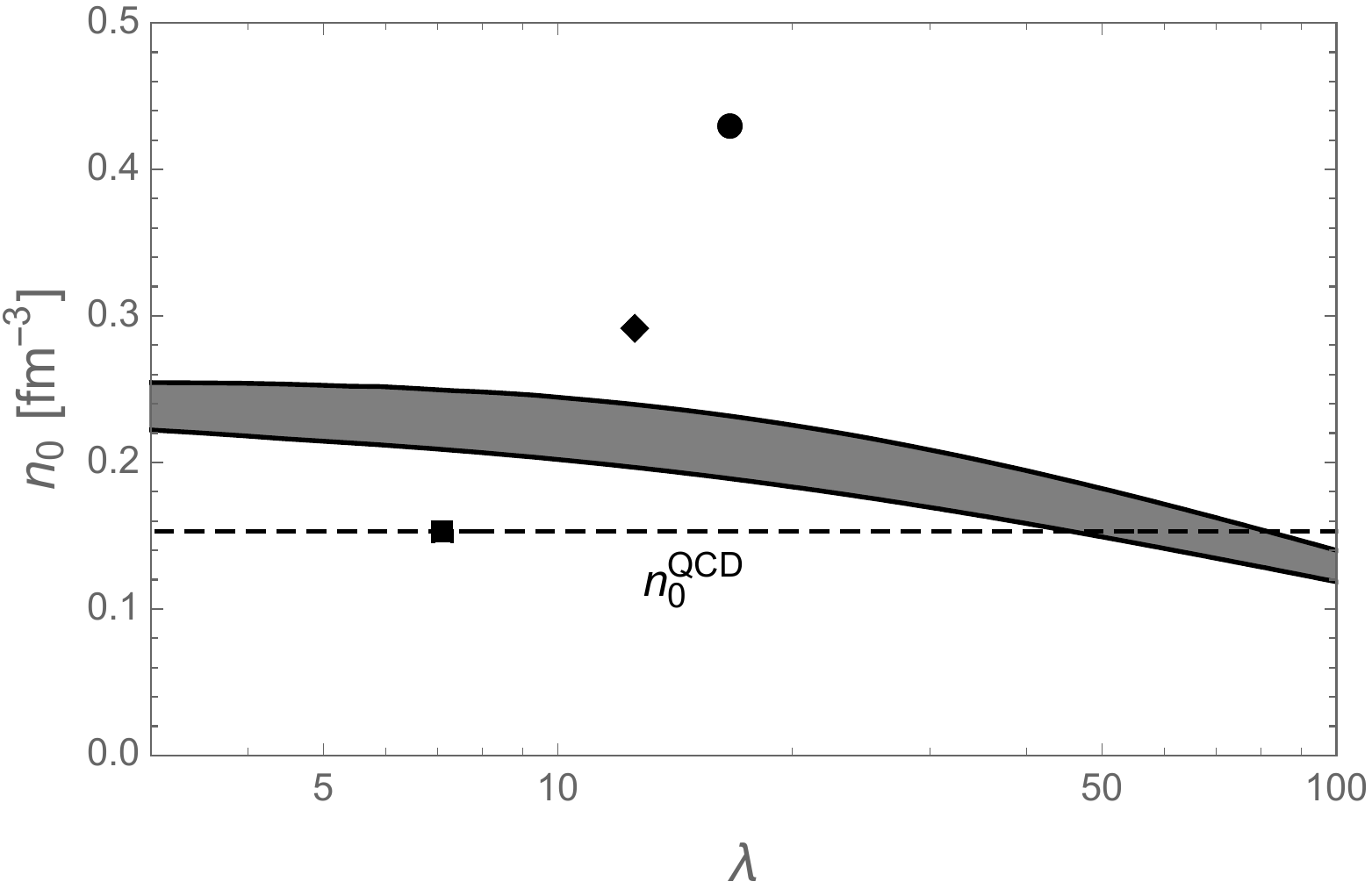}
\includegraphics[width=0.49\textwidth]{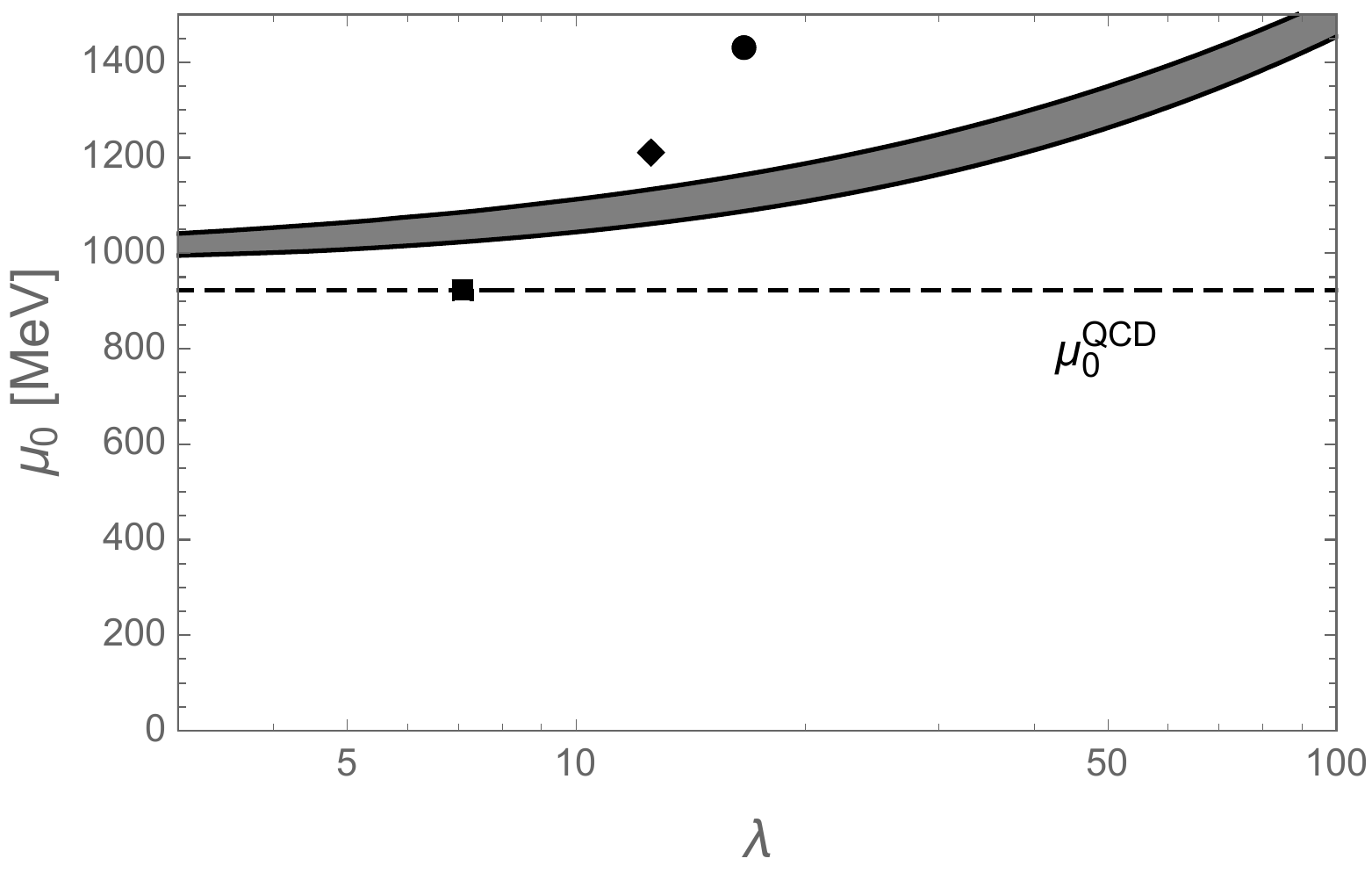}
\caption{
Saturation density of symmetric nuclear matter (left) and corresponding onset chemical potential (right) as functions of $\lambda$ for the astrophysically allowed parameter band, see left panel of Fig.\ \ref{fig:QCD1}, and for the three fits of table \ref{tab:para}. The dashed horizontal lines indicate the real-world values.}
\label{fig:QCD2}
\end{figure}

The shaded region also yields an "astrophysically allowed" range for the second model parameter $M_{\rm KK}$ because each $M_{\rm max}$ in Fig.\ \ref{fig:LR} is generated by choosing a value for $M_{\rm KK}$. Repeating this calculation for many values of $\lambda$ we can thus determine a window in the $M_{\rm KK}$-$\lambda$ plane that satisfies all astrophysical constraints. For most of the $\lambda$ range we consider, the situation is qualitatively the same as in Fig.\ \ref{fig:LR}. For very small $\lambda$, however, the scenario slightly differs: Instead of the lower bound for $R_{2.1}$, the existence of the 2.1-solar-mass star becomes the strongest constraint for the lower bound of $M_{\rm max}$; and instead of the upper bound for $\Lambda_{1.4}$, the upper bound for $R_{1.4}$ becomes the strongest constraint for the upper bound of $M_{\rm max}$.
The resulting window is the gray shaded area in the left panel of Fig.\ \ref{fig:QCD1}. For comparison, we have indicated three particular parameter choices obtained from fits to QCD vacuum properties (circle and diamond) and to saturation properties of symmetric nuclear matter (square), as explained in table \ref{tab:para}. We see that these three points do not coincide and none of them lies in the astrophysically allowed band. Having in mind that the points and the band are constructed to fit properties of vastly different systems, it is perhaps not surprising that the Witten-Sakai-Sugimoto model, at least in the simple version employed here, cannot account for all of them simultaneously. To get some further idea of the extent by which the properties of nuclear matter are violated, we have plotted the saturation density $n_0$ and the onset chemical potential $\mu_0$ for the astrophysically allowed band and the three separate fits in Fig.\ \ref{fig:QCD2}. 
If we are more modest and do not require to fit "everything" with a single parameter set but at least keep the QCD properties approximately correct, it is useful to define a "QCD window", given by the fits to the vacuum and nuclear matter: $M_{\rm KK} \simeq (949-1000)\, {\rm MeV}$ and $\lambda \simeq 7 - 17$. We have indicated this window as a red rectangle in the left panel of Fig.\ \ref{fig:QCD1}.

\begin{figure}
\centering

\includegraphics[width=0.49\textwidth]{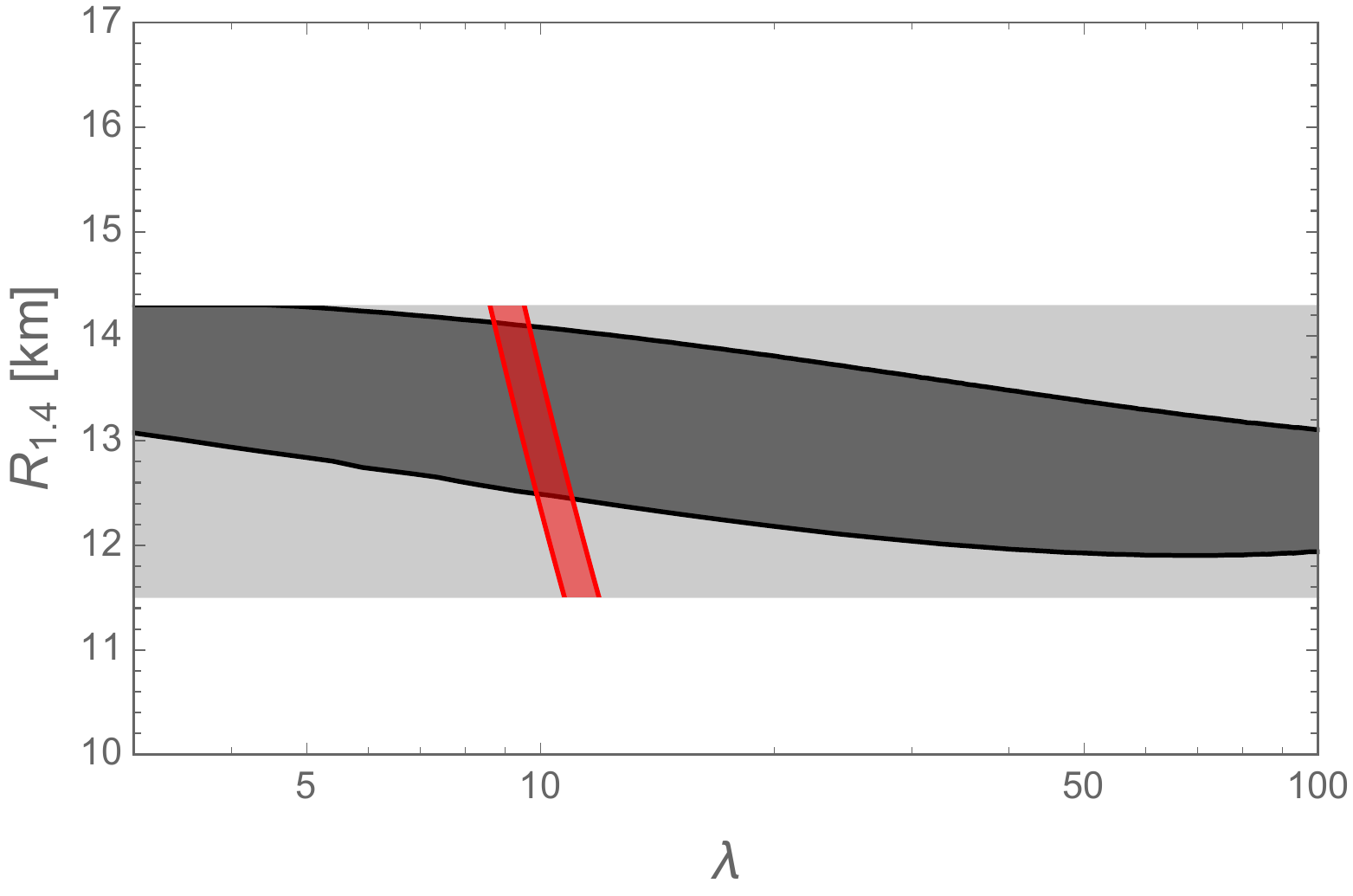}
\includegraphics[width=0.49\textwidth]{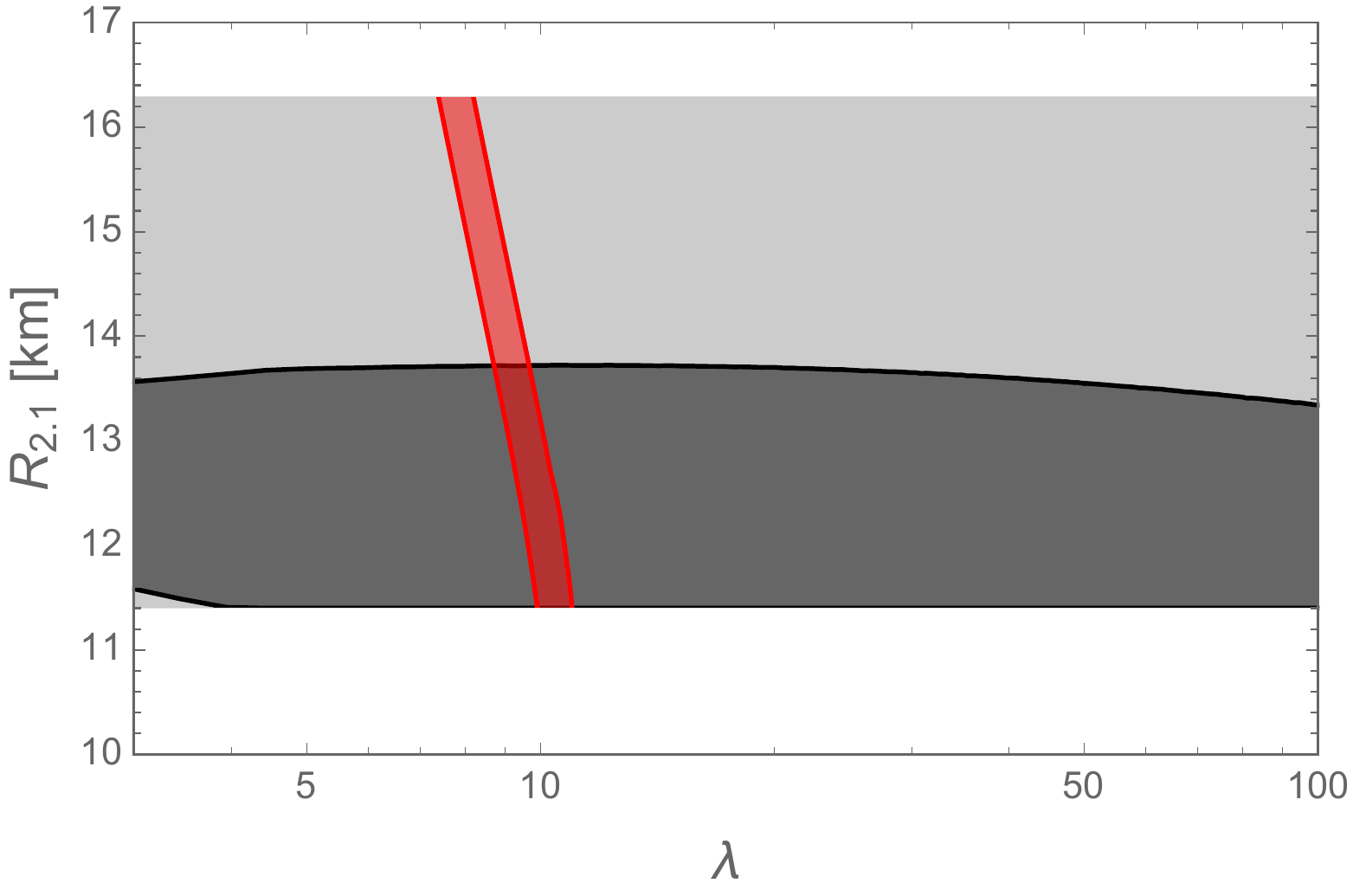}

\includegraphics[width=0.49\textwidth]{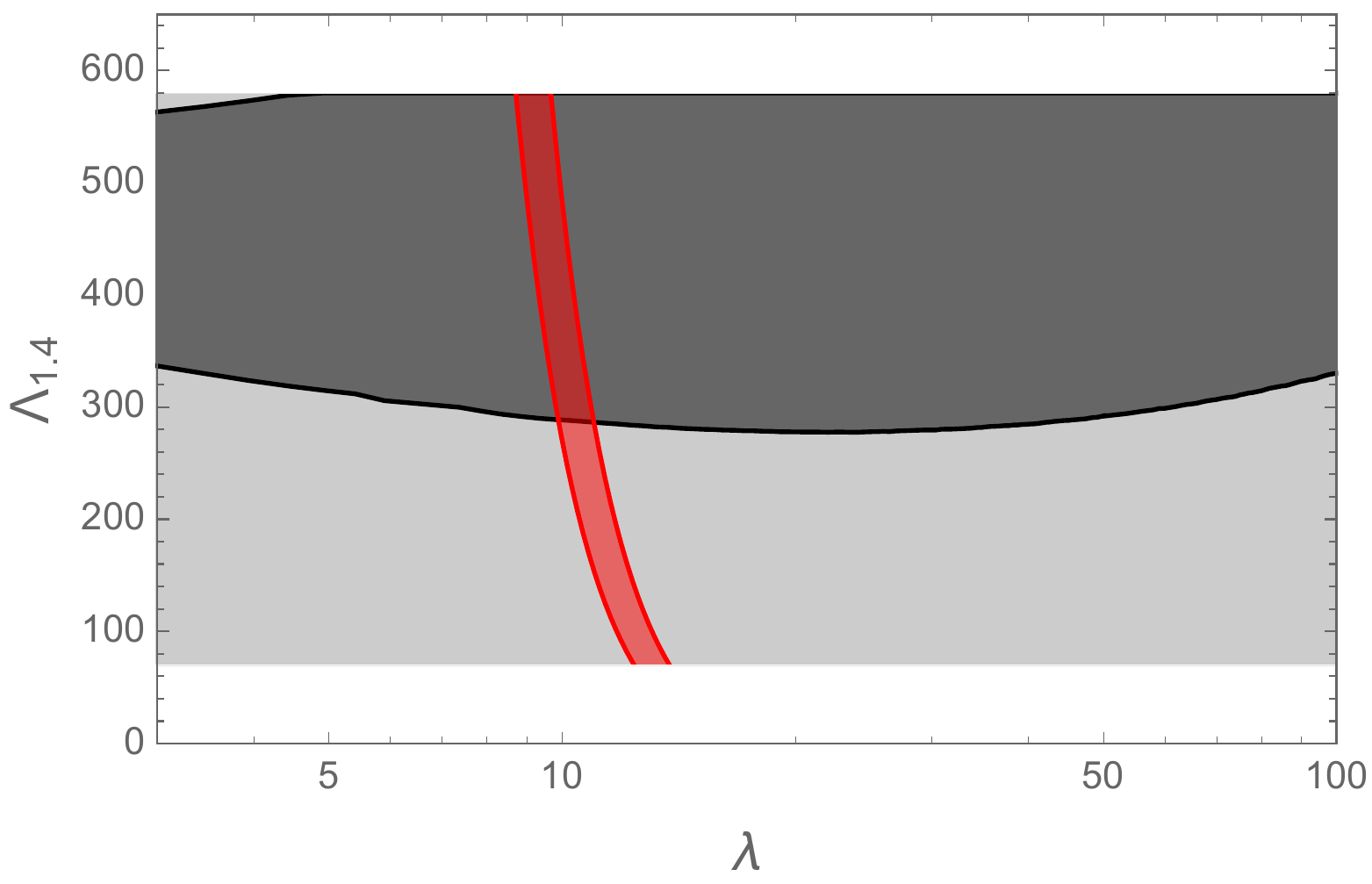}
\includegraphics[width=0.49\textwidth]{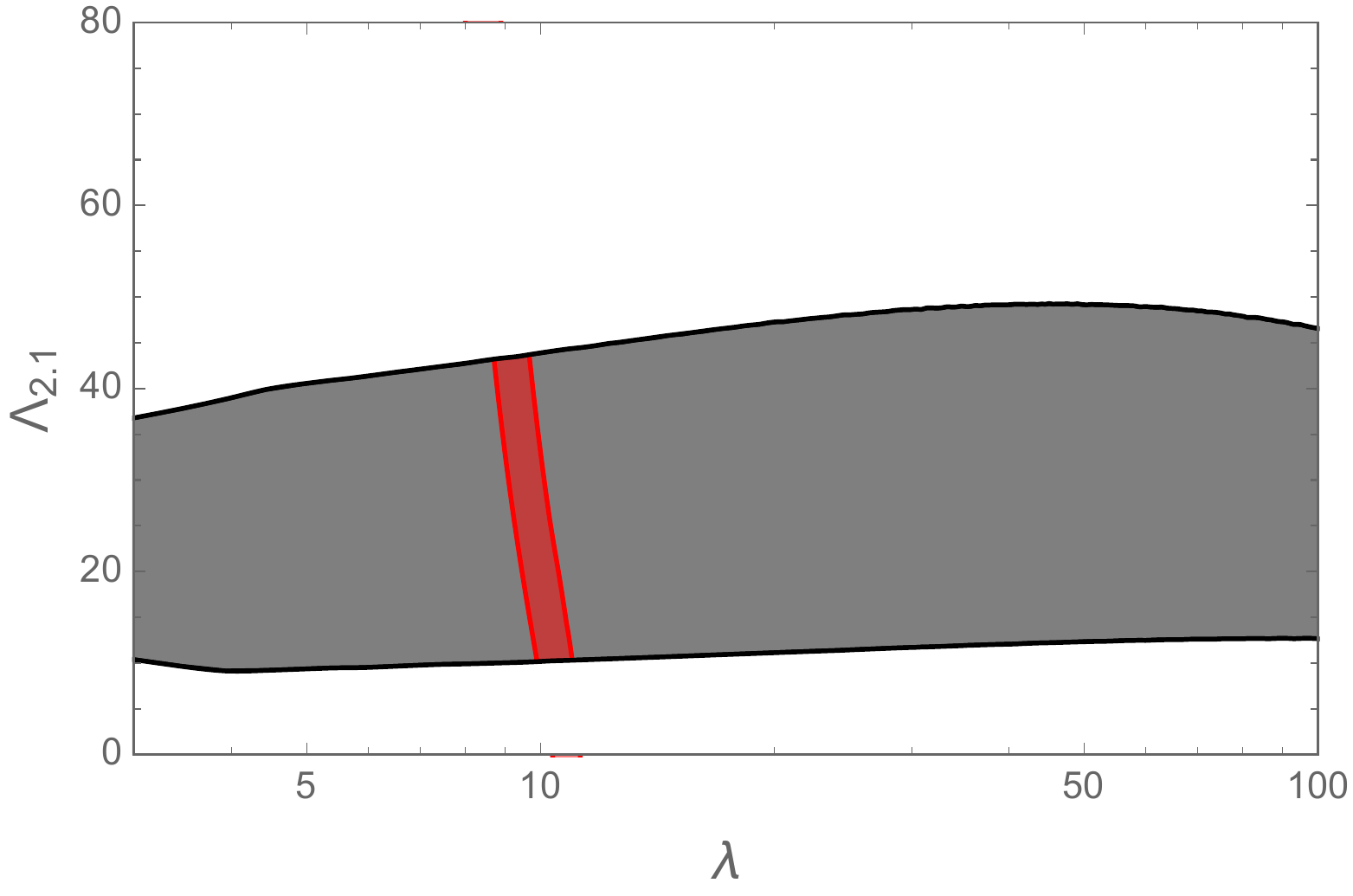}
\caption{
Same as right panel of Fig.\ \ref{fig:QCD1}, but for radius and deformability of 1.4-solar-mass and 2.1-solar-mass stars.}
\label{fig:const2}
\end{figure}

In the right panel of this figure and in Fig.\ \ref{fig:const2} we collect the constraints for all $\lambda$ obtained from the construction of Fig.\ \ref{fig:LR}. Constraints from astrophysical data alone are shown by a light gray band (obviously independent of the microscopic model parameter $\lambda$). The panel for the deformability $\Lambda_{2.1}$ does not have such a band because there is no data available from a neutron star merger with a star of that mass. The dark gray bands are the more stringent constraints obtained by combining the data with the results of the model. 
They allow us to read off predictions of the model that are completely general, i.e., with no assumptions about the model parameters $\lambda$ and $M_{\rm KK}$ (for a fixed value of the surface tension in the crust). We have collected these predictions in table \ref{tab:predict}. For the "parameter-independent" bounds we have used the upper or lower limits of the bands visible in the plots. In all cases, perhaps with the exception of $\Lambda_{2.1}$, the shapes of the bands suggest that these are the general bounds even beyond the shown $\lambda$ regime. Our predictions can further be refined by focusing on the QCD window, which is shown in each panel as a red band (cut off at the boundaries of the light gray band). The steepness of the red bands 
indicate that the observables are very sensitive to variations in $\lambda$. The refined constraints are then obtained from the intersections of the red bands with the dark gray bands (more precisely the upper or lower corner of the intersection, depending on whether 
we obtain an upper or lower limit). These values are also collected in    table \ref{tab:predict}. For instance, we find as a general prediction of the model that neutron stars cannot be heavier than 2.46 solar masses, while if we are interested to approximately reproduce vacuum and nuclear matter properties at the same time, this upper limit can be lowered to 2.40 solar masses. Similarly, for any parameter values the tidal deformability of a 1.4-solar-mass star cannot be lower than 277, the QCD window further narrows this down to a lower limit of about 286.

\section{Conclusion}
\label{sec:con}

We have employed a holographic description of zero-temperature, high-density nuclear matter and used this single, top-down formalism to construct neutron stars. In particular, since our holographic nuclear matter is allowed to become isospin asymmetric, we were able to account for electroweak equilibrium and electric charge neutrality, and we have constructed a mixed phase of nuclear matter and a lepton gas to include the crust of the star fully dynamically. We have demonstrated that the model can reproduce realistic neutron stars, and we have combined our microscopic results with the latest astrophysical data to derive constraints for mass, radius, and tidal deformability of the star.

\begin{table}[t]
    \centering
    \begin{tabular}{|c||c|c|c|c|}
    \hline
        & \multicolumn{2}{c}{parameter independent}\vline &\multicolumn{2}{c}{QCD window} \vline \\
    \hline
    & lower bound & upper bound & lower bound & upper bound \\
    \hline\hline
      $\;\;$  $M_{\rm max}\,[M_\odot]$ $\;\;$ & (2.1) & 2.46 & 2.11 & 2.40 \\
    \hline
    $R_{1.4}\, [{\rm km}]$ & 11.9 & (14.3) & 12.4& 14.1 \\
    \hline
    $R_{2.1}\, [{\rm km}]$ & (11.4) & 13.7 & (11.4) & 13.7 \\
    \hline
     $\Lambda_{1.4}$ & 277 & (580) & 286 & (580) \\
    \hline
     $\Lambda_{2.1}$ & 9.13 & 49.3 & 10.1 & 43.7 \\
    \hline
    \end{tabular}
    \caption{Constraints obtained by combining the holographic results with astrophysical data for maximal mass as well as radius and tidal deformability for 1.4-solar-mass and 2.1-solar-mass stars. Parameter-independent bounds are valid for any model parameters $\lambda$, $M_{\rm KK}$, while the QCD window defined by table \ref{tab:para} and Fig.\ \ref{fig:QCD1} gives tighter bounds. Parentheses indicate that our model does not further restrict the astrophysical data used here. 
    }
    \label{tab:predict}
\end{table}

Improvements of the holographic model are necessary for more reliable predictions, most notably a refined approximation regarding the isospin spectrum is highly desired. More straightforward improvements of the present calculation would be the construction of an inner crust as a mixed phase of pure neutron matter and nearly symmetric nuclear matter, taking into account different geometrical structures in the crust and the crust-core transition region, and computing the surface tension dynamically within the model. Other extensions are the inclusion of strangeness (and a nonzero strange quark mass), pion condensation, nonzero temperature effects, a magnetic field, and the phase transition to a chirally restored phase. Most of these ingredients have been developed already within the given model and have to be combined and possibly improved for neutron star applications. It would also be interesting to use the model to compute transport properties, as recently done in the context of dense matter within different holographic setups \cite{Hoyos:2020hmq,Hoyos:2021njg}.

\section*{Acknowledgements}
We thank N.\ Jokela and A.\ Rebhan for valuable comments. 
A.P.\ is supported by an Engineering and Physical Sciences Research Council \mbox{(EPSRC)}
Mathematical Sciences Fellowship at the University of
Southampton. N.K.\ is supported by the ERC Consolidator Grant 772408-Stringlandscape.


\bibliography{references.bib}

\nolinenumbers

\end{document}